# Propagating magnetic droplet solitons as moveable nanoscale spin-wave sources with tunable direction of emission

Morteza Mohseni,[1,*] Qi Wang,[1] Majid Mohseni,[2] Thomas Brächer,[1] Burkard Hillebrands,[1] and Philipp Pirro[1]

[1] *Fachbereich Physik and Landesforschungszentrum OPTIMAS, Technische Universität Kaiserslautern, 67663 Kaiserslautern, Germany.*

[2] *Faculty of Physics, Shahid Beheshti University, Evin, Tehran 19839, Iran*

Magnetic droplets are strongly nonlinear and localized spin-wave solitons that can be formed in current-driven nanocontacts. Here, we propose a simple way to launch droplets in an inhomogeneous nanoscopic waveguide. We use the drift motion of a droplet and show that in a system with broken translational symmetry, the droplet acquires a linear momentum and propagates. We find that the droplet velocity can be tuned via the strength of the break in symmetry and the size of the nanocontact. In addition, we demonstrate that the launched droplet can propagate up to several micrometers in a realistic system with reasonable damping. Finally, we demonstrate how an annihilating droplet delivers its momentum to a highly nonreciprocal spin-wave burst with a tunable wave vector with nanometer wavelengths. Such a propagating droplet can be used as a moveable spin-wave source in nanoscale magnonic networks. The presented method enables full control of the spin-wave emission direction, which can largely extend the freedom to design integrated magnonic circuits with a single spin-wave source.

## I. Introduction

The presence of nonlinear wave phenomena such as solitons has been widely observed and studied in many physical systems [1–7]. In magnetic materials hosting spin waves (SWs), soliton formation and manipulation are of special interest due to their potential applications in spintronic devices, for example, magnetic memories, microwave elements and neuromorphic computing systems [8–11]. Indeed, due to the presence of strong nonlinear dynamics via magnons, the quanta of SWs, magnets behave as a rich landscape for soliton formation and utilization. In addition to nonlinear phenomena, SWs have been confirmed to be suitable candidates for information transport and processing, which is due to their high group velocity and the absence of losses due to Joule heating [12–16]. In this context, understanding the mechanisms of nonlinear magnon transport opens additional degrees of freedom for new device architectures relying on wave-based computing concepts.

In order to design a universal integrated magnonic device which can be used for data storage as well as a data processing unit, it is necessary to use a system that can host solitons and propagating SWs. To fulfill this purpose, manipulation of solitons [17-25] and the excitation of nanometer SWs with a tunable wave vector are crucial tasks [13,16]. The excitation of nanometer SWs is extremely inefficient if conventional excitation mechanisms, which rely on the magnetic field generated by microwave currents, are used on the nanoscale [16]. Furthermore, the wavelength of the excited SWs is limited by the lateral size of the microwave striplines [26]. In addition, the requirement to use high-frequency currents for excitation also leads to elements with low energy efficiency. Thus, overcoming these challenging tasks requires more advanced methods, such as magnetic nano-grating [26,27], spin textures [28–30]. However, one of the main drawbacks of the presented methods so far is the lack of tunability to control the spatial direction of the SW emission.

Dynamic solitons, for example, droplets in nanomagnets, exhibit high-frequency precessional motion in the GHz range jointly with large angle changes of magnetization, promising many opportunities for the design of spintronic and high-power microwave devices [31–35]. Indeed, magnetic droplets are SW solitons that are nucleated in current-driven nanocontacts (NC) in layers with a sufficiently large perpendicular magnetic anisotropy (PMA). Therefore, using a system which can host droplets can be a suitable candidate for designing a universal magnonic element where data storage and processing are carried out in the magnonic domain. However, droplets are localized SWs that only sustain in the presence of the applied current. Thus, manipulating droplets and pushing them to propagate in an experimentally possible way remains a great challenge.

Here, we propose a simple and experimentally accessible way to launch droplets using their drift motion, thus without any electron current flowing along the droplet motion. We show that a droplet in a system with broken translational symmetry experiences a strong drift and propagates. We verify and engineer our design using micromagnetic simulations and predict that the propagating droplets acquire different momenta, depending on the strength of the broken translational symmetry of the hosting waveguides. In addition, we demonstrate that launched droplets can propagate up to several micrometers in a medium with realistic damping. The droplet propagates in a straight line without any deviation from the initial direction of motion, making droplets an attractive alternative candidate for long-range information carries in comparison to skyrmions [39]. The emission direction is normal to the direction of the broken symmetry. Finally, we show that the propagating droplet ultimately annihilates into propagating SW bursts with a well-defined momentum, which is determined by the velocity of the moving droplet. The moving droplet can thus be employed as a SW source to excite nanometer SWs. This can open up new avenues to design more advanced methods for the excitation of nanometer SWs in nano magnonic systems with controlled direction of emission. Moreover, such a system can be used as a promising design for a universal magnonic element as briefly mentioned above.

## II. Methods

Figure 1a shows the schematic design of the system, which is the free layer of a spin torque nano-oscillator (STNOs). An applied DC generates an STT, which in return nucleates the droplet beneath the NC. At one edge of the NC, the translational symmetry of the waveguide (WG) is broken. Thus, the host WG is separated into two different sections with different effective fields. These are represented by the dark and light purple areas named nucleation and propagation sections in Fig. 1a.

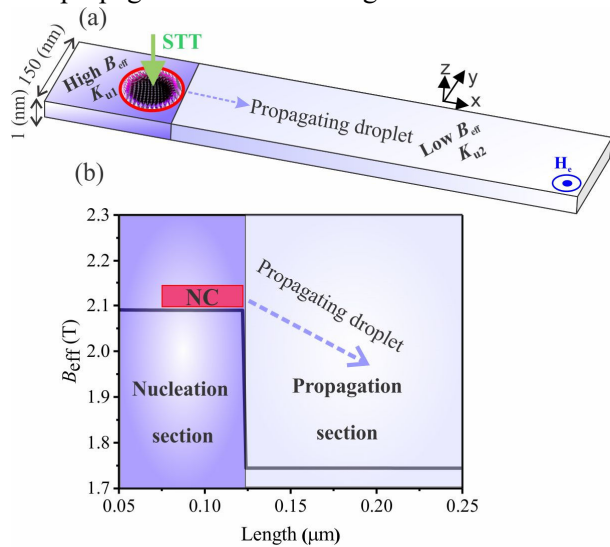

Figure 1: a) Schematic of the device design containing two sections labeled as nucleation (dark purple) and propagation sections (light purple). b) Out-of-plane component of the effective field of the system (in a relaxed state) as a function of the position along the x-axis when the external field is set to $\mu_0 H_e = 0.4$ T.

To experimentally realize such a heterostructure, the PMA or the saturation magnetization $M_s$ of the free layer must be tuned. This is possible via, for example, optical methods [40], ion radiation [41], electrical currents [42], voltage controlled magnetic anisotropy (VCMA) [43] or temperature [44]. Once the droplet is nucleated beneath the NC, it is expelled from the NC and propagates toward the section with a lower effective field as a consequence of the broken symmetry. If the DC continues to exert a torque on the magnetization of the layer, a second droplet nucleates and begins to propagate after a brief delay. This process continues as long as STT is applied, which means a train of propagating droplets can be achieved.

Micromagnetic simulations were performed using MuMax 3.0 open source software [45]. For the results presented here, as shown in Fig. 1a, we used a WG with dimensions of $4000 \times 150 \times 1$ nm$^3$ with a cell size of $2 \times 2 \times 1$ nm$^3$. The system is magnetized perpendicularly to the x-y plane with a field amplitude of $\mu_0 H_e = 0.4$ T. The parameters we used are as follows [18]: $M_s = 580$ KA/m, $A_{exch} = 15$ pJ/m, $\lambda = 1.3$ and the spin polarization $P = 0.5$. The NC diameter is set to 50 nm (unless mentioned otherwise), and the applied DC is $I = 4$ mA (along the z direction with negative polarity) in all simulations. We used a fixed uniaxial PMA strength of $K_{u2} = 600$ KJ/m$^3$ in the part with a lower effective field where the droplet propagates (propagation section).

In contrast, we vary the uniaxial PMA of the nucleation section $K_{u1}$ to discuss the impact of the energy step on the droplet momentum. For example, Fig. 1b shows the out-of-plane component of the effective field for $K_{u1} = 700$ KJ/m$^3$, corresponding to a difference of $\Delta B_{eff} = 16.7\%$ between the effective fields of the two sections. For simplicity, we only apply a single STT pulse with a duration of $t_{pulse} = 0.5$ ns to nucleate only one droplet.

### III. Results and discussions

#### A. Droplet in a quasi-lossless medium

First, we set the damping of the system close to zero at the end of the DC pulse to study the features of a propagating droplet in a quasi-lossless medium.

Figure 2a shows a sequence of snapshots of the droplet propagation for $K_{u1} = 700$ KJ/m$^3$. The 0.5 ns current pulse is applied at $t = 0$ ns (as depicted in Fig. 2c). Once the droplet is nucleated, it drifts toward the section with a lower effective field, distinguished via a vertical red line near the NC edge. Since the damping is set to a negligible value, the droplet continues to propagate to infinite distance (compared to the size of the modeled system) without a change in its size, evidencing energy conservation [38].

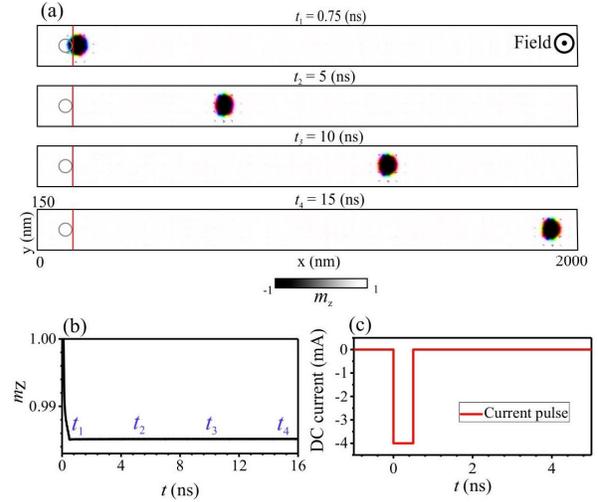

Figure 2: Droplet propagation in a quasi-lossless medium for $K_{u1} = 700$ KJ/m$^3$; a) Time sequence snapshots of the propagating droplet; b) Normalized out-of-plane ($m_z$) component of the magnetization of the waveguide during the motion of the droplet; c) The shape of the DC pulse applied to the nanocontact. Gray circles in (a) show the nanocontact.

This is also visible in Fig. 2b, which shows the integrated $m_z$ component of the magnetization in the WG as a function of time. Once the droplet is nucleated, the value of $m_z$ remains constant throughout the entire propagation distance. As evident from Fig. 2a, the droplet propagates in a straight line without a change in the direction of its propagation.

In order to explore the droplet dynamics during its motion, in Fig. 3a, we present zoomed-in snapshots of the droplet under motion. Indeed, the lower effective field on the right side of the NC acts as a trap since it allows the effective dipolar moment of the droplet to acquire an in-plane component. Therefore, a spatial variation of the precessional phase appears, which finally breaks the coherency of the precession at the droplet boundaries. Under this condition, the droplet experiences a strong and complex shape deformation via an interplay between the anisotropy and exchange energies. This finally leads to a frequency change and a strong linewidth enhancement of the propagating droplet.

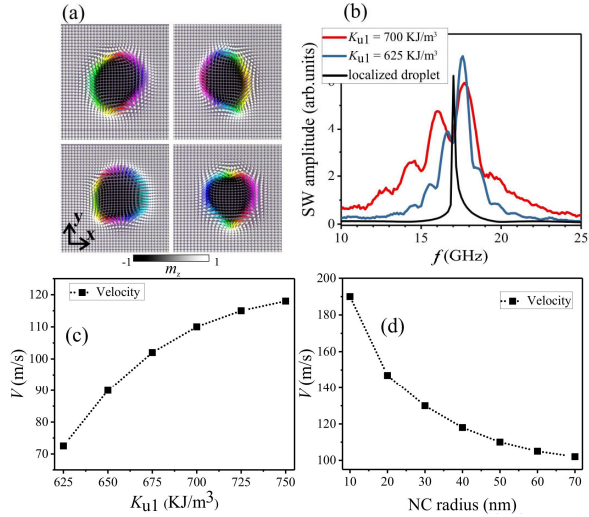

Figure 3: a) Snapshots of the propagating droplet. Arrows indicate the direction of the magnetic moments; b) Frequency spectra of the localized and propagating droplets; c) Velocity of the propagating droplet versus the PMA strength of the nucleation region; d) Velocity of the droplet as a function of the nanocontact diameter.

The frequency spectra of the droplet for $K_{u1} = 700$ KJ/m$^3$ (red curve), $K_{u1} = 625$ KJ/m$^3$ (blue curve) and for a localized droplet in the propagation section (black curve) are presented in Fig. 3b. In addition to the main droplet frequency, sidebands appear evidencing the droplet's inertial effects, i.e., its tendency to resist the exerted torque due to the broken symmetry [48]. Moreover, it is evident that a higher $K_{u1}$ leads to a larger linewidth in the frequency spectra of the moving droplet. In fact, an increase in the value of $K_{u1}$ increases the effective field beneath the NC. This leads to a higher frequency of the (localized) droplet before propagation and consequently a higher storage of the energy in the droplet's precessional boundaries. This higher precessional frequency leads to a faster spatial deformation of the droplet structure during motion. This means that the propagating droplet experiences faster shape deformations. Furthermore, this is accompanied by higher propagation speeds.

The motion of a dynamic particle-like droplet can also transform the energy stored in the precessional motion into the effective kinetic energy of the translational motion [36–38]. Similar to Newton's law $\overrightarrow{dX}/dt = \vec{V}$, the droplet continues to move with a constant velocity due to its inertia. For this particular case, which is presented in Fig. 2a, the velocity equals $V = 110$ m/s. In general, the velocity depends on the strength of the broken translational symmetry or, in other words, the difference between the effective fields of the two regions. This is visible in Fig. 3c, where the droplet velocity is shown as a function of the strength of the PMA in the nucleation section. It is evident that an increase in the value of $K_{u1}$ increases the effective field beneath the NC. The larger step enables the droplet to acquire a higher momentum and, accordingly, a higher velocity when it launches.

It is known that the frequency of the droplet can be tuned via the size of the NC such that smaller NCs lead to higher frequencies [34]. We set $K_{u1} = 700$ KJ/m$^3$, and we use the same conditions for the droplet nucleation; however, the size of the NC diameter is varied to show how it influences the initial droplet momentum. Fig. 3d demonstrates that the velocity of the droplet increases by decreasing the size of the NC. In addition to the given explanations, one should expect smaller droplets to carry smaller effective masses due to their size and higher frequencies [36]. Therefore, in the presence of the same torque (effective field step), they acquire higher speeds.

We now analyze the dependence of the droplet frequency $f$ (and its closest sidebands, which are labeled as $f^+$ and $f^-$) on its velocity, as depicted in Fig. 4a. We used a Lorentzian fitting function to find the given frequencies. Evidently, the propagating droplets with higher velocities exhibit higher frequencies (black curve in Fig. 4a). This is based on the fact that the droplets with higher frequencies should feature smaller effective masses and, consequently, higher propagation speeds; see Ref [36-38].

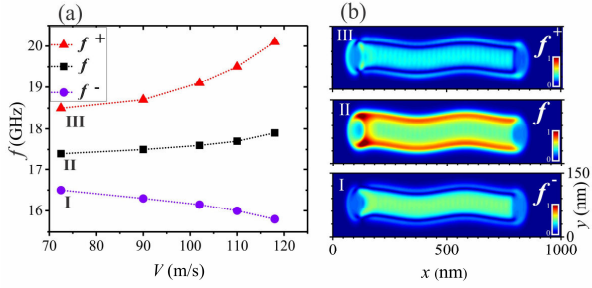

Figure 4: a) Dependence of the propagating droplet frequency (and the closest sidebands) on the droplet velocity; b) Spatial profile of the main droplet mode ($f$) and the two nearest sidebands ($f^+$ and $f^-$) during propagation labeled as I, II and III in (a).

The spatial distribution of the eigenmodes of the propagating droplet is of special interest for applications in information transport via magnons, which are presented in Fig. 4b. The main droplet mode labeled as $f$ (II in Fig. 4b) features a higher amplitude at the boundaries of the trajectory of the propagating droplet. This is caused by the strong shape deformations of the droplet, as discussed before. However, the energy of the sidebands is mainly distributed at the center of the trajectory of the moving droplet (labeled as I and III in Fig. 4b). This is in accordance with our expectations since such modes for a localized droplet feature a chiral profile due to the interplay between the STT and the induced force [48]. It is interesting to note that the shape deformations of the droplet during propagation lead to a small modulation of its trajectory (toward the $y$-axis) along the propagation direction.

### B. Droplet in a dissipative medium

Thus far, we discussed droplet propagation in a lossless WG. However, in reality, magnetic materials exhibit internal losses manifesting themselves as Gilbert damping. In the following, we fix the damping to realistic values, and we will address how the damping influences a propagating droplet.

First, we set $K_{u1} = 700$ KJ/m$^3$ and α = 0.01 with the same nucleation conditions (0.5 ns DC pulse) as discussed previously. Snapshots of the simulated system over the droplet lifetime are presented in Fig. 5a. Once the droplet has launched, it decreases in size since its energy dissipates due to damping. In addition, as its size decreases throughout the propagation, its velocity increases. Indeed, damping reduces the effective mass of the droplet rapidly by losing energy and, consequently, accelerates very quickly [36,38,47]. Finally, the droplet annihilates by exciting a SW burst.

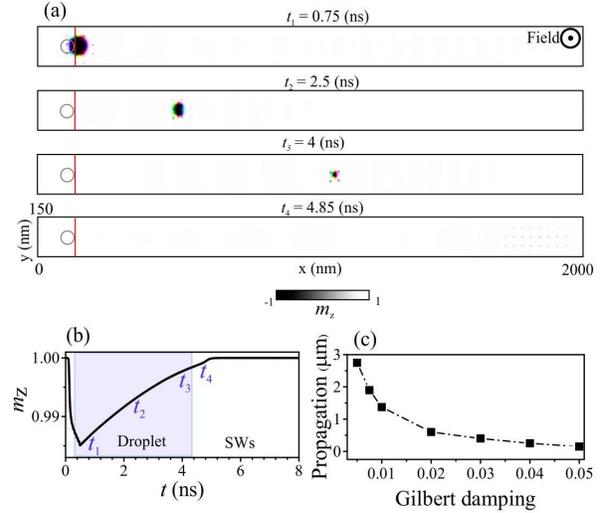

Figure 5: Droplet propagation in a dissipative medium (α = 0.01) when the $K_{u1} = 700$ KJ/m$^3$; a) Sequence of snapshots of the propagating droplet. The droplet finally annihilates to a SW burst; b) Normalized integrated out-of-plane ($m_z$) component of the magnetization of the waveguide under the motion of the droplet; c) Propagation distance of the droplet before annihilation as a function of the Gilbert damping factor.

The shrinking of the droplet is also visible in Fig. 5b, which shows the time trace of the $m_z$ component of the system. Once the droplet is nucleated and the DC is set to zero, $m_z$ begins to increase, and finally, the droplet annihilates. The distance travelled by the droplet before it annihilates is affected by the damping rate. In principle, the losses have a reverse impact on the lifetime of the soliton. Figure 5c shows the distance at which a droplet propagates before its annihilation as a function of the damping rate. In fact, increasing the damping increases the rate of energy loss of the droplet. Therefore, the droplets sustain shorter distances in the presence of higher damping. However, the propagation distance is still remarkable (several hundreds of nm), even if the magnetic system exhibits a relatively significant damping rate of α = 0.05. In addition, it is worth mentioning that the damping of the system in the presence of a droplet can be modified via spin currents [49].

We now investigate the droplet dynamics by analyzing the frequency spectra of the propagating droplet at different time intervals.

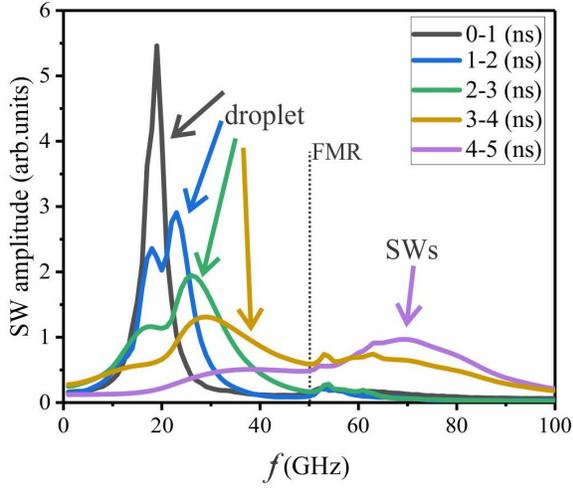

Figure 6: Frequency spectrum of the propagating droplet in a dissipative medium at different time intervals. FMR refers to the ferromagnetic resonance.

As shown in Fig. 6, two mechanisms influence droplet motion. The amplitude of the propagating droplet decays with time because its energy dissipates due to damping. More interestingly, the frequency of the droplet increases as it accelerates during propagation, similar to what has been discussed above. Furthermore, due to the acceleration of the droplet during its motion, the linewidth becomes larger, as discussed in the context of Fig. 3b.

Indeed, the linewidth of the propagating droplet increases continuously until its frequency is high enough to (partially) overlap with the eigenmodes of the magnon band upon annihilation (after 3 ns in Fig. 6). Thus, as we have observed before, the droplet decays resonantly to a burst of SWs when it annihilates.

In the following, we will analyze the characteristics of SWs emitted by the annihilation of the droplet. For this purpose, we set the damping to 0.01 and again vary the PMA of the nucleation region. In Fig. 3c, we showed that a droplet acquires a higher initial momentum by changing the strength of the PMA in the nucleation region. In the presence of damping, the droplet additionally experiences acceleration, which makes it difficult to determine the instantaneous velocity from the simulation. Nevertheless, it is possible to define an averaged velocity of the droplet $V_{ave}$ as the total distance it propagates divided by the time of its propagation, given by the time span between emission and annihilation. Figure 7a shows this averaged velocity $V_{ave}$ as a function of the PMA strength. In the presence of damping, the droplet shows a higher $V_{ave}$ and a longer propagation distance as it acquires a higher momentum in the presence of a larger effective field step. However, annihilating droplets with different averaged velocities demonstrate different behaviors in terms of their SW excitation.

The insets of Fig. 7a show snapshots of the droplets just before their annihilation. It is evident that the size of the soliton before its annihilation depends on its initial momentum. Namely, the annihilating droplet is larger if it had a smaller initial momentum in the presence of the same damping.

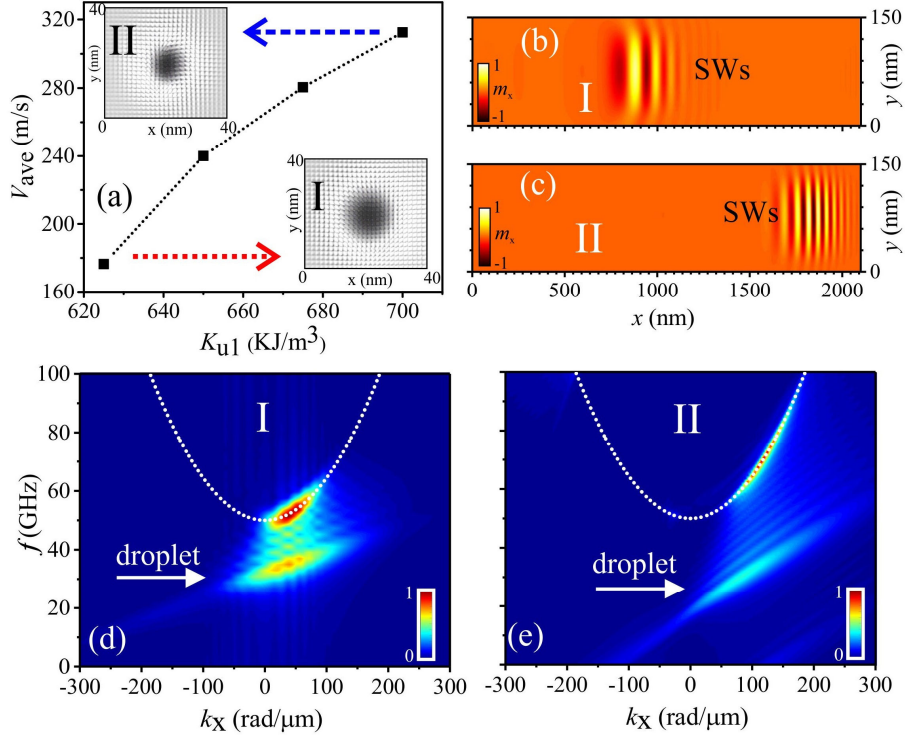

Figure 7: a) Averaged velocity $V_{ave}$ of the droplet as a function of the PMA strength of the nucleation region. Insets show the droplet size for different PMA (marked by arrows) at the end of their lifetime; b) Snapshots of the spin-wave bursts of the annihilating droplet, corresponding to I and II; d-e) Magnon band and the nonreciprocal population of the spin waves caused by the annihilating droplets. Dashed white lines are the spin-wave dispersions in the waveguide.

Interestingly, these different droplet scales lead to different SW bursts upon annihilation. This can be seen more clearly in Fig. 7b-c, which show snapshots of the SW bursts when these two different droplets annihilate. As shown in the figure, the larger droplet with lower $V_{ave}$ decays to SWs with larger wavelengths.

To gain further insight into how the droplet momentum is delivered to the SWs, Figs. 7d-e show the FFT in space and time of the droplet when it decays into the SW burst. Moreover, from these Fourier spectra, it is possible to infer how the annihilating soliton populates the magnon band. As briefly discussed above, the frequency and linewidth of the droplet increase during propagation until it partially overlaps with the SW eigenmodes of the WG. More interestingly, it is evident that the droplet with smaller velocity (larger annihilation size and smaller momentum) leads to the excitation of SWs with smaller wave vectors and, conversely, the droplet with higher velocity (smaller annihilation size and larger momentum) excites SWs with larger wave vectors. Thus, the maximum wave vector available for SW excitation is in direct correspondence with the momentum of the annihilating droplet. An additional important fact can also be seen from Fig. 6: The propagating droplet only annihilates toward the direction of its propagation. This is reflected in the fact that the dispersion curve of the SWs shows the existence of a highly nonreciprocal magnon population. These results demonstrate that the momentum of the propagating soliton is directly transferred to propagating nonsolitonic SWs.

### C. Comparison to other spin torque based spin-wave emitters

The designed system uses propagating magnetic droplet solitons to excite nanometer SWs. The most important advantage of the presented method is the spatial mobility of the SW source. Using this method can greatly extend the freedom to design magnonic integrated devices with a single SW emitter which provides full tunablity of the emission direction to feed the entire circuit. This can help to simplify the design and reduce the fabrication procedures and relevant costs significantly which can serve to make magnonics more compatible with CMOS. Indeed, full control of the emission direction in a single device is completely lacking in similar methods based on inductive, spin transfer torque- and spin orbit torque (SOT)- driven propagating SWs [16, 50-55]. This high nonreciprocity of the SW excitation is very suitable for integrated magnonics since the excitation of counter propagating waves which are not fully avoidable using other methods, can disturb the functionality of the devices for logical operations [13,56]. In addition, control of the SW wave vector, which is a necessary ingredient for data processing using SWs is achieved by controlling the speed of the propagating droplet via tuning the strength of the broken symmetry. In contrary to the STT and SOT based devices in which higher currents are required to excite shorter SWs [54, 55], using droplet for the same purpose is independent of the dc amplitude injected into the NC. Thus, not only it seems as a more energy efficient way to control the wave vector of the SWs, but it gives an additional degree of freedom to tune them. Needless to mention that the droplet itself can store and transport information. Therefore, such a system can help to design a universal magnonic element for data storage and processing on a single chip.

## IV. Conclusions

In conclusion, we have demonstrated a simple and feasible system for launching droplet solitons into magnetic WGs. The launched droplet propagates to infinite distances with a constant velocity in a lossless medium without changing its size or frequency. However, in the presence of losses, the droplet accelerates while reducing in size by losing energy. Even in the presence of substantial damping, the droplet can propagate up to several hundreds of nanometers. Upon annihilation, the droplet delivers its momentum to nonsolitonic SWs by annihilating into a nonreciprocal SW burst. The momentum of the SW bursts can be tuned by the momentum of the propagating droplet. Such a moving droplet can be considered a mobile SW source with a full control of the emission direction in nanoscale magnonic networks. Using this method may open up new routes toward more flexible mechanisms of the excitation of nanometer SWs for magnonic computation platforms, and can generate new ideas to realize a universal magnonic devices.

This project is funded by the Deutsche Forschungsgemeinschaft (DFG, German Research Foundation) - TRR 173 - 268565370 ("Spin+X", Project B01), the Nachwuchsring of the TU Kaiserslautern, the DFG Priority Programme "SPP2137 Skyrmionics" and the European Research Council Starting Grant 678309 MagnonCircuits. Majid Mohseni acknowledges support from the Iran Science Elites Federation (ISEF) and the Iran National Elites Foundation (INEF). Valuable discussions with Vasyl Tyberkevych is appreciated.

-----------------------------------------------------------

*Correspondence: mohseni@rhrk.uni-kl.de